\documentclass[12pt]{iopart}
\usepackage{epsf}
\usepackage{graphicx}

\begin{document}

\title[The influence of dipole-dipole interaction on  entanglement...]{The influence of dipole-dipole interaction on  entanglement in nondegenerate two-photon Tavis-Cummings model with atomic coherence}

\author{Bashkirov E.K., Mastuygin M.S.}

\address{Samara State University, 1 academician Pavlov Str., Samara, 443011, Russia }
\ead{mast12basket@rambler.ru}
\begin{abstract}

The entanglement of two dipole-coupled atoms with nondegenerate two-photon transitions interacting with two-mode field in lossless cavity has been investigated.
It shows that the entanglement is dependent on the initial atomic states.  The possibility of considerable  growth of atomic entanglement due atomic coherence and dipole-dipole interaction is shown in the case of great mean values of  thermal photons.

\end{abstract}

\vspace{2pc}
\maketitle

\section{Introduction}
Entanglement is a key resource which distinguishes quantum from classical information theory. It plays a central role
in quantum information, quantum computation and communication,  and quantum cryptography  \cite{Nielsen}. In order to function optimally these applications require
 maximally entangled states.  However, interactions with the environment always occur, and will degrade the quality
 of the entanglement. Although the interaction between the environment and quantum systems can lead to decoherence,
 it can also induce  entanglement \cite{Plenio}. Thus, understanding and investigating entanglement of mixed states becomes one of the actual  problem of quantum information. Recently,  Bose et al.  \cite{Bose} have shown that entanglement can always arise in the interaction of an arbitrary large system in any
 mixed state with a single qubit in a pure  state,  and illustrated this using the Jaynes-Cummings interaction of a
 two-level atom in a pure state with a field in a thermal state at an
 arbitrary high temperature.   Kim et al. \cite{Kim} have investigated the atom-atom entanglement in the system of
  two identical   two-level atoms with one-photon  transition induced by a single-mode
 thermal field. They showed that a chaotic field with minimal information can entangled atoms which were
 prepared initially in a separable
 state.  Zhou et al. \cite{Zhou2} have considered the same problem for nonidentical atoms with different couplings. The entanglement between
 two identical  two-level atoms through  nonlinear two-photon interaction with one-mode thermal field has been
 studied by Zhou  et al. \cite{Zhou1}. They  showed that atom-atom entanglement induced by nonlinear interaction is larger than that
 induced by linear interaction. In \cite{Bash1} has discovered that two atoms can be entangled also through nonlinear  nondegenerate two-photon interaction with two-mode thermal field. The influence of dipole-dipole interaction on entanglement between two cubits induced by one-mode and two-mode thermal field has been investigated in \cite{Aguiar}-\cite{Bash2}.

 The problem of creating or controlling the atomic entanglement is greatly related to the atomic coherence of population between different levels. Hu et al \cite{Hu1},\cite{Hu2} have shown that the entanglement between two atoms induced by one-mode thermal field can be manipulated by changing the initial parameters of the atoms, such as the superposition coefficients and the relative phases of the initial atomic coherent state and the mean photon number of the cavity field.
 In \cite{Hu3} Hu and Fang study  the effect of the atomic coherence on the entanglement of two two-level atoms interacting with two-mode thermal fields through a nondegenerate two-photon process, which is a nonlinear interaction  They have  found that for some atomic initial state the
 entanglement induced by nonlinear interaction may be larger than that induced by linear interaction. They have also discovered that
 entanglement may be greatly enhanced due to atomic coherence. This provides a method by which one can control the entanglement of the whole system. But the influence of the dipole-dipole interaction on the coherence-enhanced entanglement has not be considered.

 In this paper we study the   effect of the atomic coherence on two atoms  entanglement induced by two-mode thermal fields taking into account the dipole-dipole interaction.

\section{The model}
In this model, we consider a system composed of two identical two-level atoms coupled via dipole-dipole interaction. The frequency of the atomic transition
for atoms is $\omega_0$. Both atoms are also interacting with two modes of quantum electromagnetic field with frequencies $\omega_1$ and $\omega_2$
through a nondegenerate two-photon process. For simplicity we ignore the Stark shift and assume the lossless cavity modes are at the two-photon
resonance with the atomic transition, i.e. the condition $\omega_1+\omega_2=\omega_0$
takes place. Then, the Hamiltonian of the considered system in
interaction picture can be written in the following form:
$$
H_I = \hbar g \sum\limits_{i=1}^2 (a_1^{+}a_2^+ R_i^- + R_i^+ a_1 a_2 )
 + \hbar \Omega (R_1^+ R_2^- + R_2^+ R_1^- ),\eqno{(1)}
$$
where $a_j^+$ and $a_j$ are the creation and the annihilation operators of photons of  $j$th
 cavity mode ($j=1,2)$, $R_i^{+}$ and $R_i^{-}$ are the raising and the lowering
operators for the $i$th atom ($i=1,2$),  $g$ is the coupling constant
between atom and the cavity field, $\Omega$ is the coupling
constant of the dipole interaction between the atoms and $|+\rangle$ and $|-\rangle$ are the excited and the ground states of
a single two-level atom. The two-atom wave function
can be expressed as a combination of state vectors of the form $|\it v_1,\it v_2\rangle = |\it v_1|| \it v_2\rangle$, where $\it v_1, v_2 =+,-.$

The density operator for the atom-field system follows a unitary time evolution generated by the evolution operator
$U_I(t) =\exp[-\imath H_I t/\hbar].$ On the two-atom
basis, $|+,+\rangle,\> |+,-\rangle,\> |-,+\rangle,\>  |-,-\rangle$,  the analytical form of
the evolution operator $U_I(t)$  is given by
\begin{center} $\quad$
$$
U(t) = \left (
\begin{array}{cccc}\vspace{2mm}
U_{11} & U_{12} & U_{13} & U_{14}\\ \vspace{2mm}
U_{21} & U_{22} & U_{23} & U_{24}\\ \vspace{2mm}
U_{31} & U_{32} & U_{33} & U_{34}\\ \vspace{2mm}
U_{41} & U_{42} & U_{43} & U_{44}\\
\end{array} \right ).\eqno{(2)}$$
\end{center}
where $$U_{11} = 1 + 2 a_1 a_2
\frac{A}{\lambda} a_1^+ a_2^+,\quad U_{14} =  2 a_1 a_2
\frac{A}{\lambda} a_1 a_2,\quad
U_{41} =  2 a_1^+ a_2^+ \frac{A}{\lambda} a^+_1 a^+_2,
$$
$$
U_{44} = 1 + 2 a_1^+ a_2^+ \frac{A}{\lambda} a_1 a_2,\quad
U_{12} = U_{13} = a_1 a_2 \frac{B}{\theta},\quad U_{21} = U_{31} =
\frac{B}{\theta} a_1^+ a_2^+, $$
$$U_{24}=U_{34} =
\frac{B}{\theta} a_1 a_2 ,\quad U_{42} = U_{43} = a_1^+ a_2^+
\frac{B}{\theta},$$
$$
U_{22} = U_{33} = $$ $$ = \frac{\exp\left [- \imath
\frac{g}{2}(\alpha+\theta)t \right]}{4\theta} \left \{ [1 -
\exp(\imath g \theta t)] \alpha + 2 \theta
\exp(\imath \frac{g}{2}(3 \alpha + \theta) t] + \theta [1 +
\exp(\imath g \theta t)]\right \},
$$\vspace{2mm}
$$
U_{23} = U_{32}= $$ $$ = \frac{\exp\left [- \imath
\frac{g}{2}(\alpha+\theta)t \right]}{4\theta} \left \{ [1 -
\exp(\imath g \theta t)] \alpha -  2 \theta
\exp(\imath \frac{g}{2}(3 \alpha + \theta) t] + \theta [1 +
\exp(\imath g \theta t)]\right \},
$$
and
 $$ A = \exp \left [ - \imath \frac{g\alpha}{2}  t\right ] \left
 \{ \cos \left (\frac{g\theta}{2} t \right ) + \imath
 \frac{\alpha}{\theta} \sin \left ( \frac{g\theta}{2} t \right )
 \right \} - 1, $$\vspace{2mm}
$$ B = \exp \left [ - \imath \frac{g}{2}(\alpha + \theta)  t \right ] \left
 [1 - \exp(\imath g \theta t)\right ], $$\vspace{2mm}
$$ \alpha = \frac{\Omega}{g},\quad \lambda = 2(a_1 a_2 a_1^+ a_2^+ + a_1^+ a_2^+ a_1
a_2), \quad \theta = \sqrt{8(a_1 a_2 a_1^+ a_2^+ + a_1^+ a_2^+ a_1
a_2)+ \alpha^2}.
$$
The initial cavity mode state are assumed to be the
thermal two-mode state
$$\rho_F(0)= \sum\limits_{n_1} \sum\limits_{n_2} p_1(n_1)p_2(n_2) |n_1,n_2\rangle \langle n_1,n_2|. \eqno{(3)}$$
The weight functions are
$$p_i(n_i)= \frac{{\bar n_i}^{n_i}}{(1+{\bar n_i})^{n_i+1}},$$
where ${\bar n_i}$  is the mean photon number in the $i$th cavity
mode, ${\bar n_i} = (\exp[\hbar \omega_i/k_BT]-1]^{-1}$, $k_B$ is
the Boltzmann constant and $T$ is the equilibrium cavity temperature.

We consider also the initial state of each atoms to be prepared in a coherent superposition of the two levels, that is,
$$ |\Psi_1(0)\rangle  = \cos \theta_1 |+\rangle + e^{\imath \varphi_1} \sin\theta_1 |-\rangle,\quad
 |\Psi_2(0)\rangle  = \cos \theta_2 |+\rangle + e^{\imath \varphi_2} \sin\theta_2 |-\rangle.$$
Here  $\theta_1$ and $\theta_2$ denote the amplitudes of the polarized atoms, and $\varphi_1$ and $\varphi_2$ are relative phases of the two atoms, respectively. So the initial density matrix for the two atoms can be written as
\begin{center} $\quad$
$$
\rho_A(0) = \left (
\begin{array}{cccc}\vspace{2mm}
\rho_{11}(0) & \rho_{12}(0) & \rho_{13}(0) & \rho_{14}(0)\\ \vspace{2mm}
\rho_{12}^*(0) & \rho_{22}(0) & \rho_{23}(0) & \rho_{24}(0)\\ \vspace{2mm}
\rho_{13}^*(0) & \rho_{23}^*(0) & \rho_{33}(0) & \rho_{34}(0)\\ \vspace{2mm}
\rho_{14}^*(0) & \rho_{24}^*(0) & \rho_{34}^*(0) & \rho_{44}(0)\\
\end{array} \right ),\eqno{(4)}$$
\end{center}
where the matrix elements are expressed as follows:
$$\rho_{11}(0) = \cos^2\theta_1 \cos^2\theta_2, \quad \rho_{12}(0) = \cos^2\theta_1 \cos\theta_2 \sin\theta_2 e^{-\imath \varphi_2},$$
$$\rho_{13}(0) = \cos\theta_1 \sin\theta_1 \cos^2\theta_2 e^{-\imath \varphi_1}, \quad
 \rho_{14}(0) = \cos \theta_1 \sin\theta_1 \cos\theta_2 \sin\theta_2 e^{-\imath(\varphi_1+ \varphi_2)},$$ $$  \rho_{22}(0) = \cos^2\theta_1 \sin^2\theta_2,\quad
 \rho_{23}(0) = \cos \theta_1 \sin\theta_1 \cos\theta_2 \sin\theta_2 e^{-\imath(\varphi_1- \varphi_2)},$$ $$\rho_{33}(0) = \sin^2\theta_1 \cos^2\theta_2,
\rho_{24}(0) = \cos\theta_1 \sin\theta_1 \sin^2\theta_2 e^{-\imath \varphi_1},\quad \rho_{34}(0) = \sin^2\theta_1  \cos\theta_2 \sin\theta_2e^{-\imath \varphi_2},$$ $$
\rho_{44}(0)= 1 - \rho_{11}(0)-\rho_{22}(0)-\rho_{33}(0).$$

\section{Results}

To investigate the entanglement between atoms one can obtain the time-dependent reduced atomic density
operator by tracing the combined atom-field density operator over the field variables:
$$\rho_A(t) = Tr_F U(t) \rho_F(0) \otimes \rho_A(0) U^+(t).\eqno{(5)}$$
For the two-qubit system described by density operator $\rho_A(t)$, a measure of entanglement can be defined in
terms of the negative eigenvalues $\mu_i^-$ of partial transpose \cite{Peres},\cite{Horod} of the reduced density matrix
$$\varepsilon= - 2 \sum\limits \mu_i^-.$$
 When $\varepsilon= 0$  two atoms are
separable and $\varepsilon > 0$ means the atom-atom entanglement.
The case $\varepsilon= 1$ indicates maximum entanglement.

Substituting Eqs (2)-(4) into Eq. (5) one can obtain the atomic density matrix at time t, accordingly, we can write down the partial transpose matrix
\begin{center} $\quad$
$$
\rho_A(t) = \left (
\begin{array}{cccc}\vspace{2mm}
\rho_{11} & \rho_{12} & \rho_{13} & \rho_{14}\\ \vspace{2mm}
\rho_{12}^* & \rho_{22} & \rho_{23} & \rho_{24}\\ \vspace{2mm}
\rho_{13}^* & \rho_{23}^* & \rho_{33} & \rho_{34}\\ \vspace{2mm}
\rho_{14}^* & \rho_{24}^* & \rho_{34}^* & \rho_{44}\\
\end{array} \right ),\eqno{(6)}$$
\end{center}
the evident expressions of the matrix elements are given in Appendix.

In Fig. 1 we plot the negativity as a function of $gt$ for a fixed value of dipole strength  $\alpha = 0.1$ and   weak thermal two-mode field with $\bar n_1 = \bar n_2 =0.01$. The Fig 1a) corresponds to atoms in the  incoherent states and Fig.2b) corresponds to them  in two different coherent  states. Comparing Fig. 1 with
that for atoms  without dipole interaction \cite{Hu3}, we find that they are clearly different. Ignoring the dipole interaction we can obtain that  the atom-atom entanglement is greatly enhanced owing to the atomic coherence.  On the contrary the  presence of an atomic  coherence for dipole coupled atoms leads to reduction of the entanglement maximum value.  In Fig. 2 we plot the negativity as a function of $gt$ for a fixed value of dipole strength  $\alpha = 0.1$ and   weak thermal two-mode field with $\bar n_1 = \bar n_2 =0.2$ but different values of relative phase $\Delta\varphi =\varphi_1-\varphi_2 $. The atoms  are assumed to be prepared in coherent states with $\theta_1= \theta_2 =\pi/4$. Clearly, the entanglement may be stronger for appropriate value of relative phase $\Delta\varphi$.
In Fig. 3 and Fig. 4 we show the entanglement time dependent for intensive two-mode thermal field. With increase of the mean photon numbers (see Fig. 1 - Fig. 4) the value of atom-atom negativity decreases. But comparing Fig. 3 and Fig. 4  with the same in \cite{Hu3} one can easily obtain that in view of dipole interaction we gain essentially large enhancement of entanglement degree transferring from incoherent to coherent atomic initial state. The results indicate that we can produce more strong entanglement by initial atomic coherence and dipole-dipole interaction.

\begin{figure}[!h]
\begin{tabular}{cc}
\mbox{a)} & \mbox{b)} \\
\includegraphics[scale=0.55]{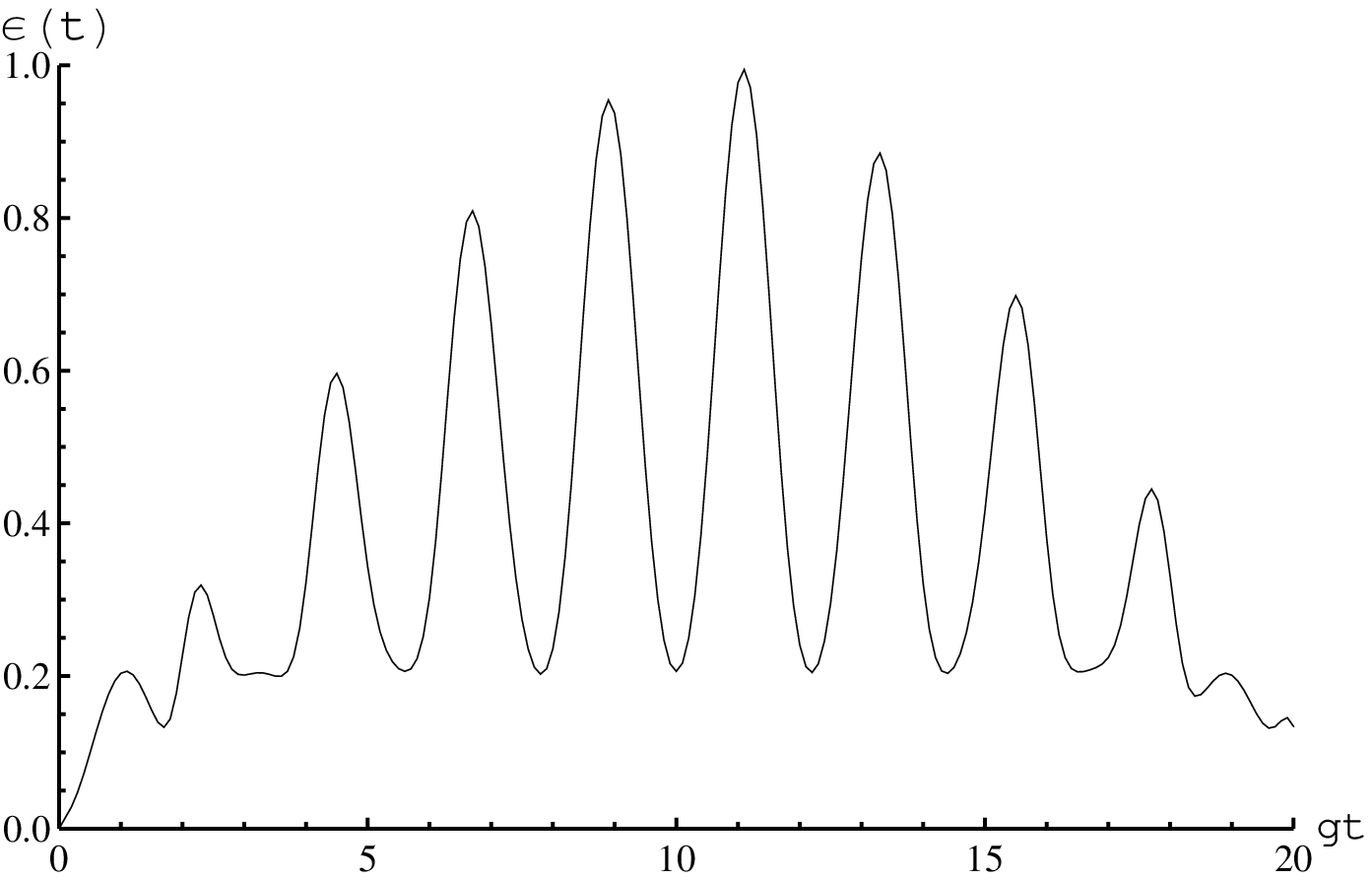} & \includegraphics[scale=0.55]{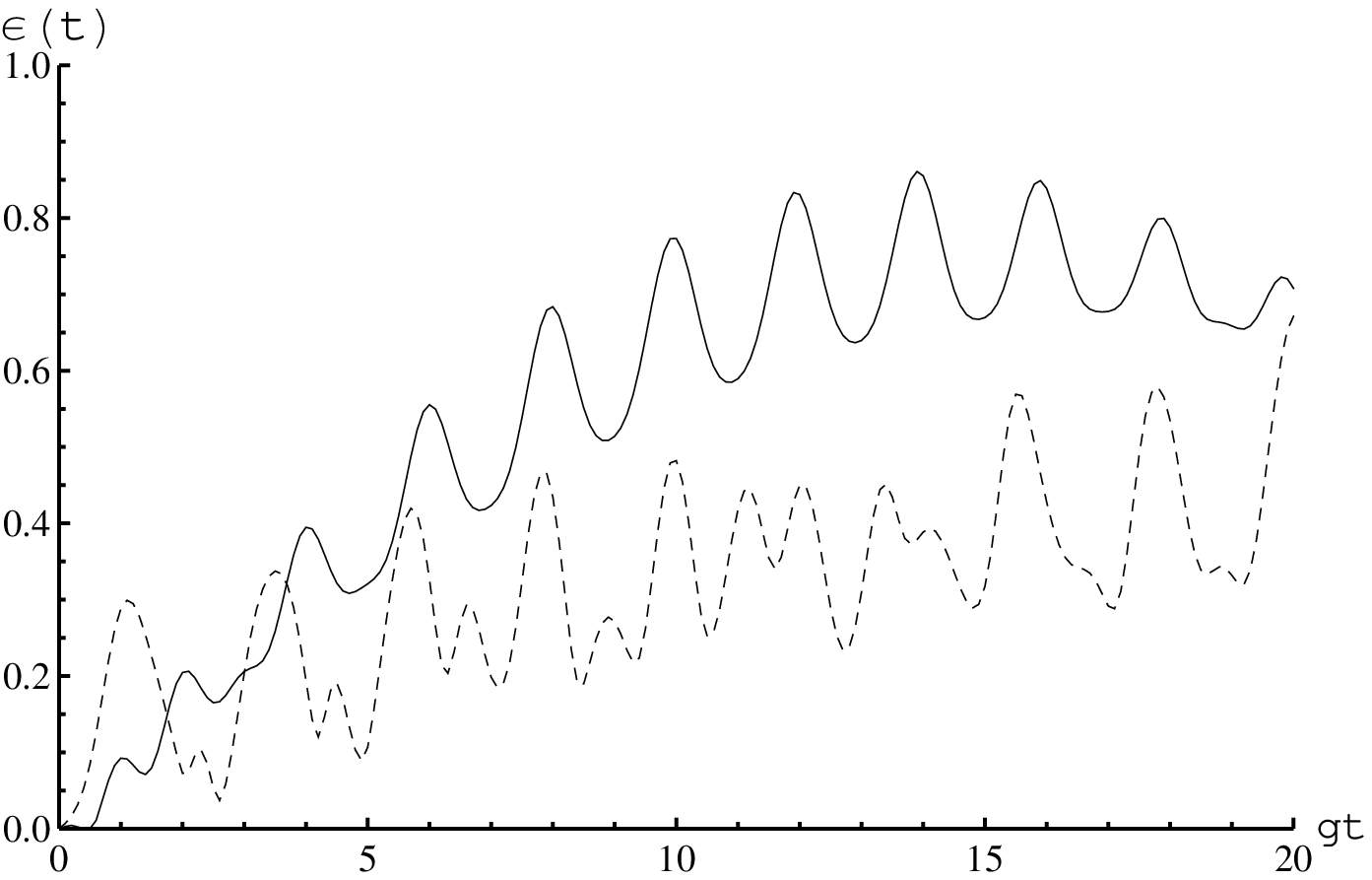}
\end{tabular}
\caption{The negativity as a function of $gt$  for the model  with  $\bar n_1 = \bar n_2 =0.01$ and  $\alpha = 0.1$:  a)   $|\Psi(0)\rangle _1= |+\rangle, |\Psi(0)\rangle_2 = |-\rangle$, b) $|\Psi(0)\rangle_1=(1/\sqrt{2}(|+\rangle + |-\rangle), |\Psi(0)\rangle_2 = (1/\sqrt{2}(|+\rangle - |-\rangle) $ (solid) and
$|\Psi(0)\rangle_1=(1/\sqrt{2}(|+\rangle + |-\rangle), |\Psi(0)\rangle_2 = (1/\sqrt{2}(|+\rangle + |-\rangle) $ (dashed). }
\end{figure}
\begin{figure}[!h]
\begin{tabular}{cc}
\mbox{a)} & \mbox{b)} \\
\includegraphics[scale=0.55]{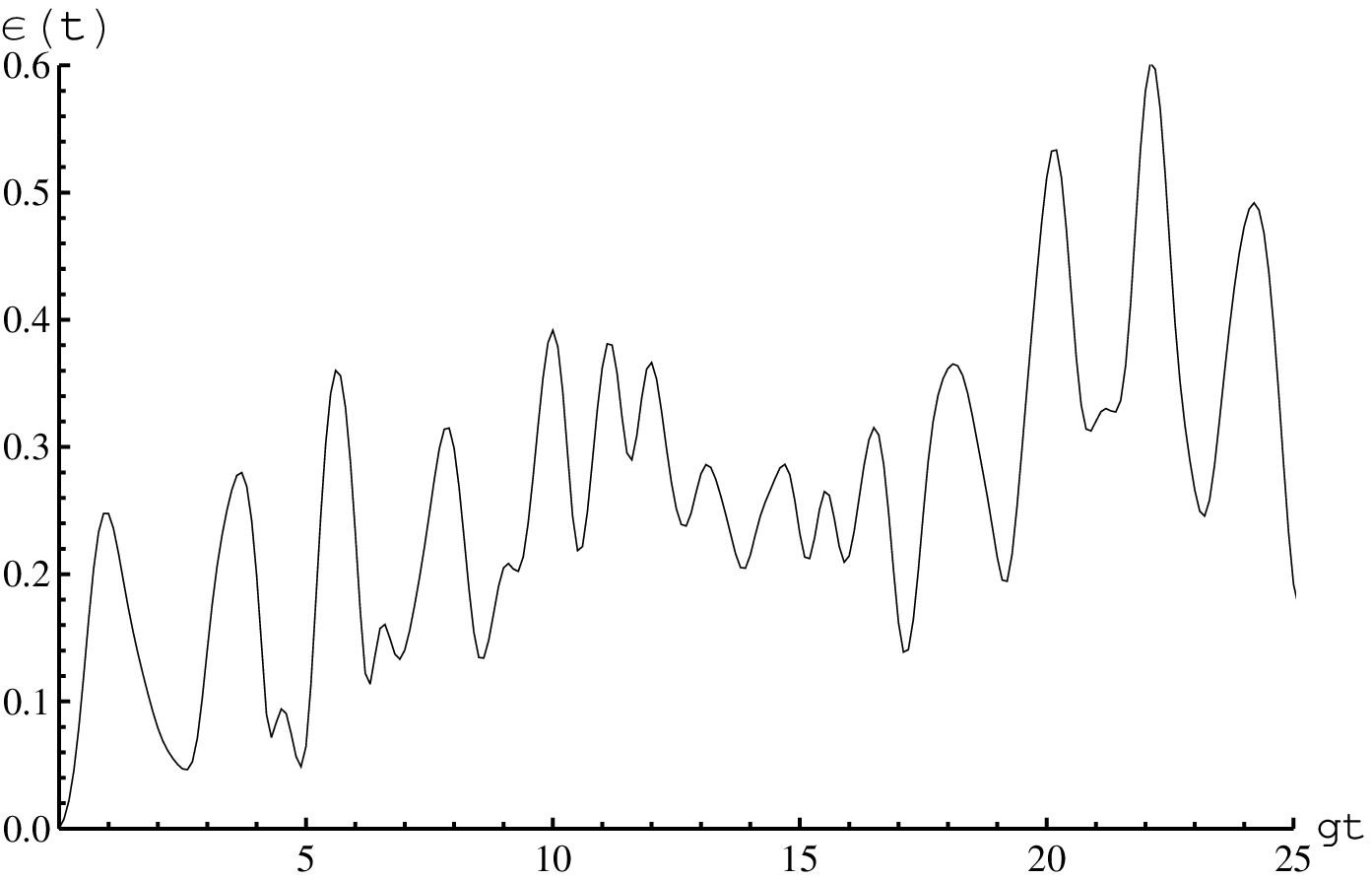} & \includegraphics[scale=0.55]{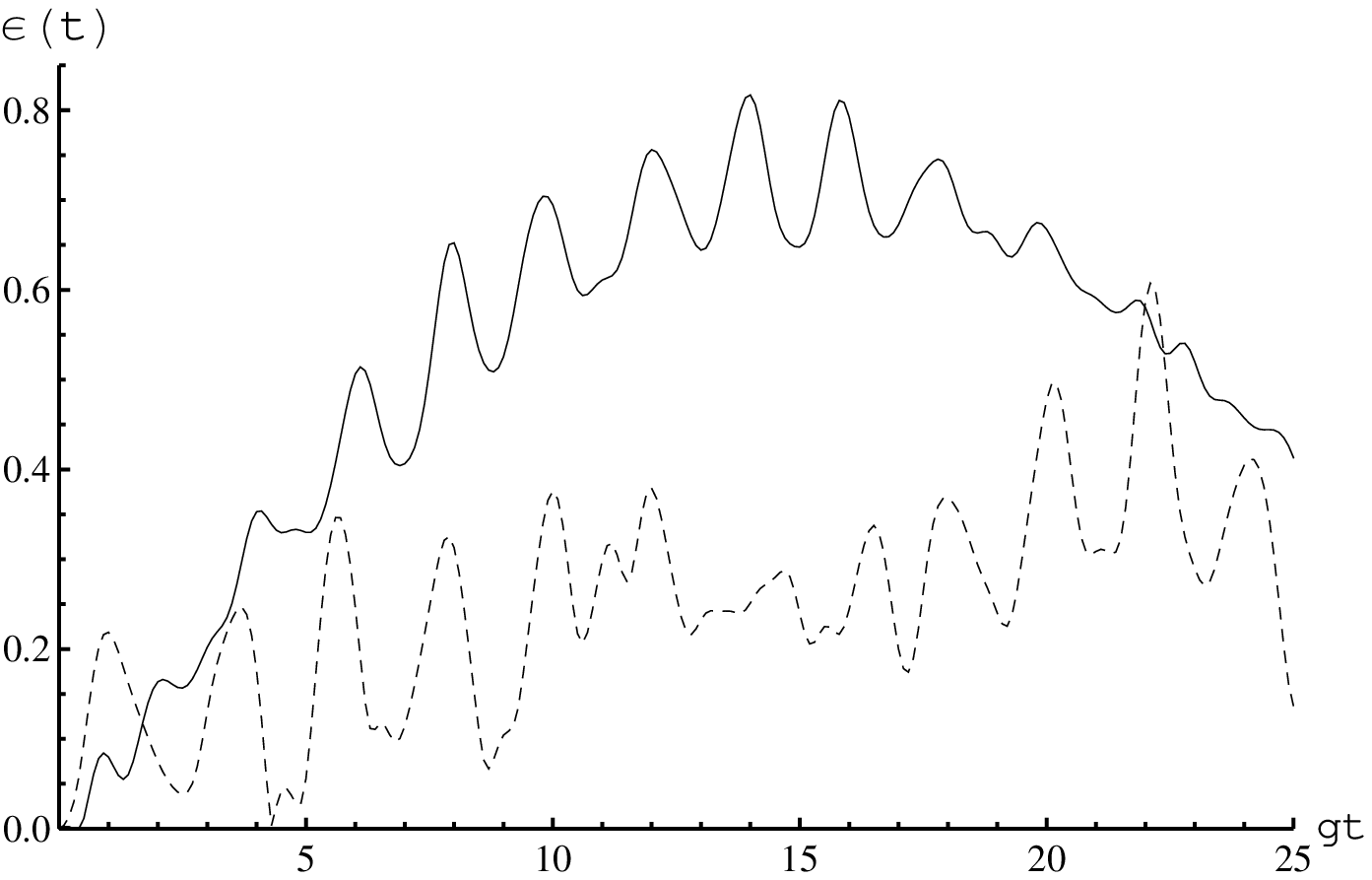}
\end{tabular}
\caption{The negativity as a function of $gt$  for the model  with  $\bar n_1 = \bar n_2 =0.2$ and  $\alpha = 0.1$,  and atoms to be prepared in coherent states
 $|\Psi(0)\rangle_1=(1/\sqrt{2}(|+\rangle + |-\rangle), |\Psi(0)\rangle_2 = (1/\sqrt{2}(|+\rangle + |-\rangle) $:  a)   $\Delta\varphi =0$, b)  $\Delta\varphi =\pi$ (solid) and  $\Delta\varphi =\pi/6$ (dashed). }
\end{figure}

\begin{figure}[!h]
\begin{tabular}{cc}
\mbox{a)} & \mbox{b)} \\
\includegraphics[scale=0.55]{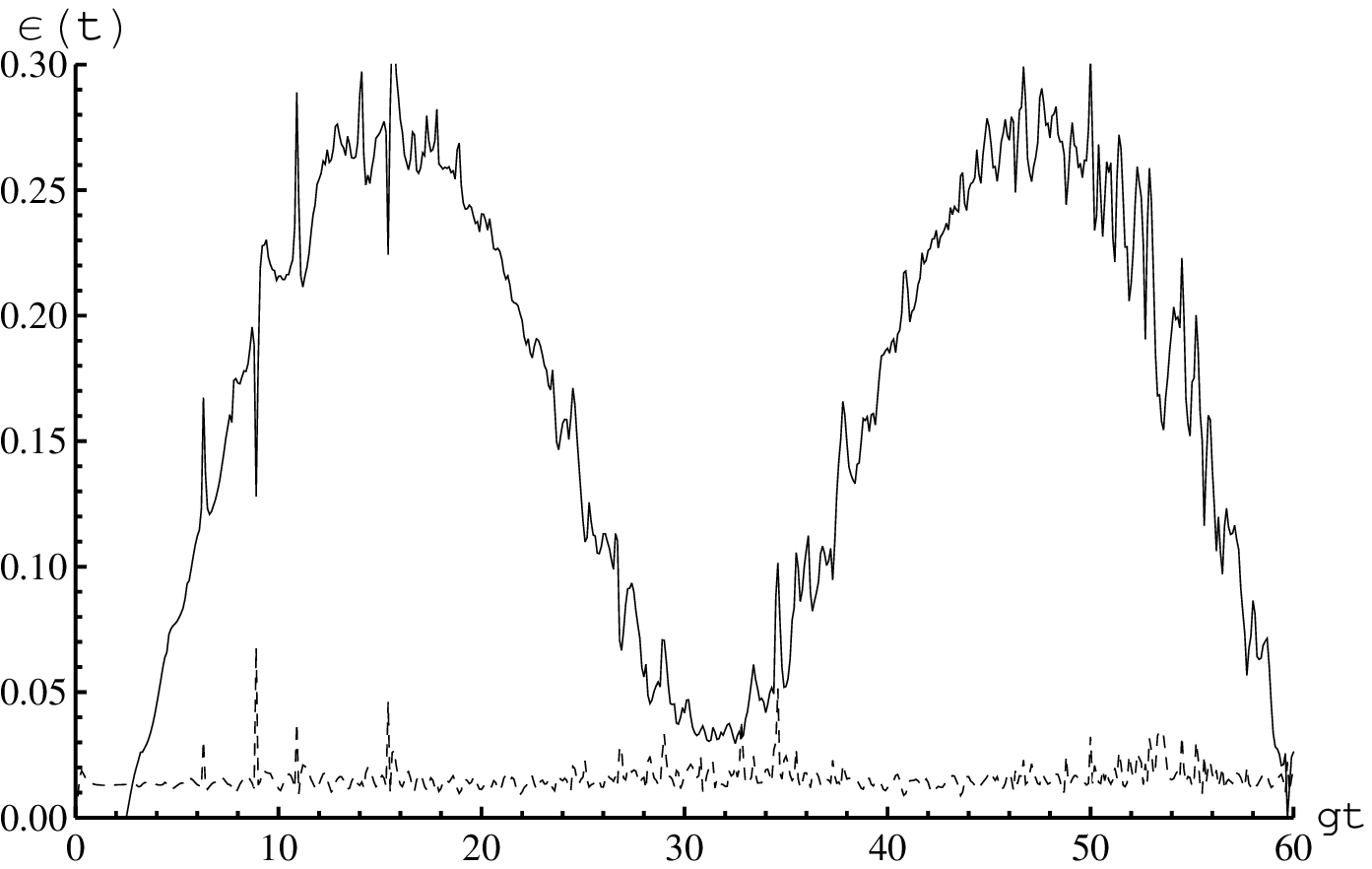} & \includegraphics[scale=0.55]{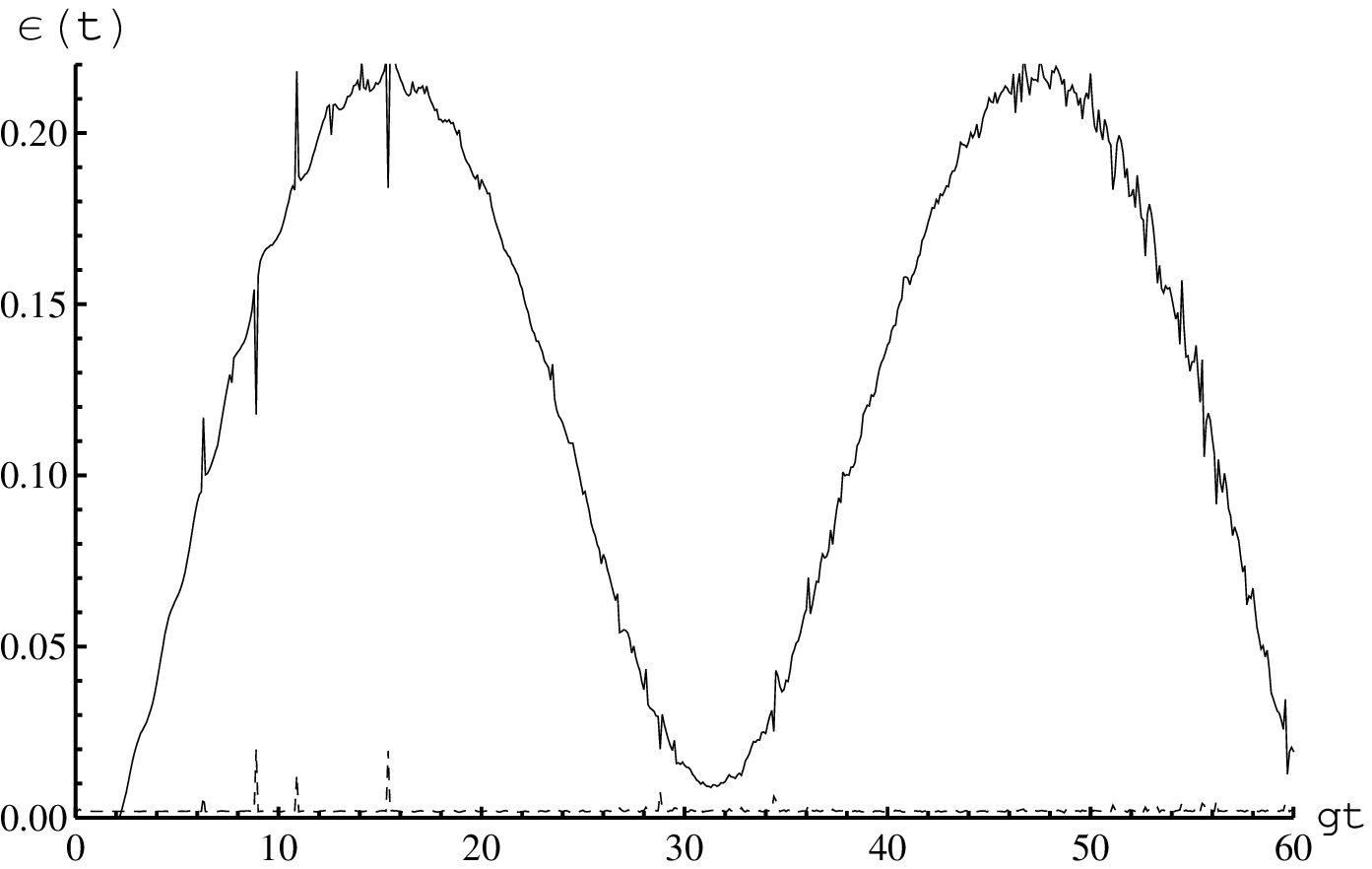}
\end{tabular}
\caption{The negativity as a function of $gt$  for the model  with    $\alpha = 0.1$:  a) $\bar n_1 = \bar n_2 =10$,  $|\Psi(0)\rangle _1= |+\rangle, |\Psi(0)\rangle_2 = |-\rangle$ (dashed), $|\Psi(0)\rangle_1=(1/\sqrt{2}(|+\rangle + |-\rangle), |\Psi(0)\rangle_2 = (1/\sqrt{2}(|+\rangle - |-\rangle) $ (solid), b) $\bar n_1 = \bar n_2 =40$,  $|\Psi(0)\rangle _1= |+\rangle, |\Psi(0)\rangle_2 = |-\rangle$ (dashed), $|\Psi(0)\rangle_1=(1/\sqrt{2}(|+\rangle + |-\rangle), |\Psi(0)\rangle_2 = (1/\sqrt{2}(|+\rangle - |-\rangle) $ (solid). }
\end{figure}
\begin{figure}[!h]
\begin{tabular}{cc}
\mbox{a)} & \mbox{ b)} \\
\includegraphics[scale=0.55]{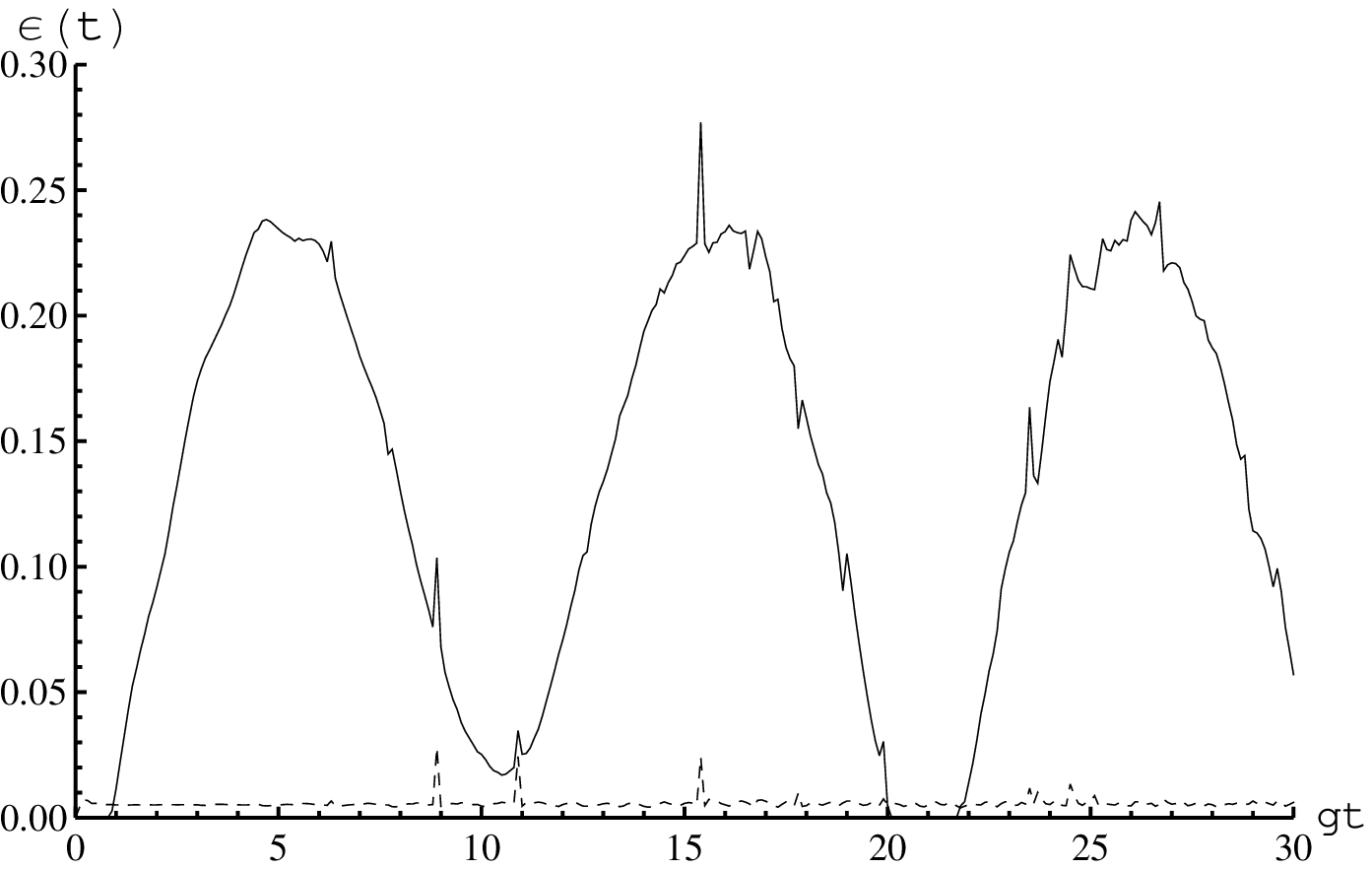} & \includegraphics[scale=0.55]{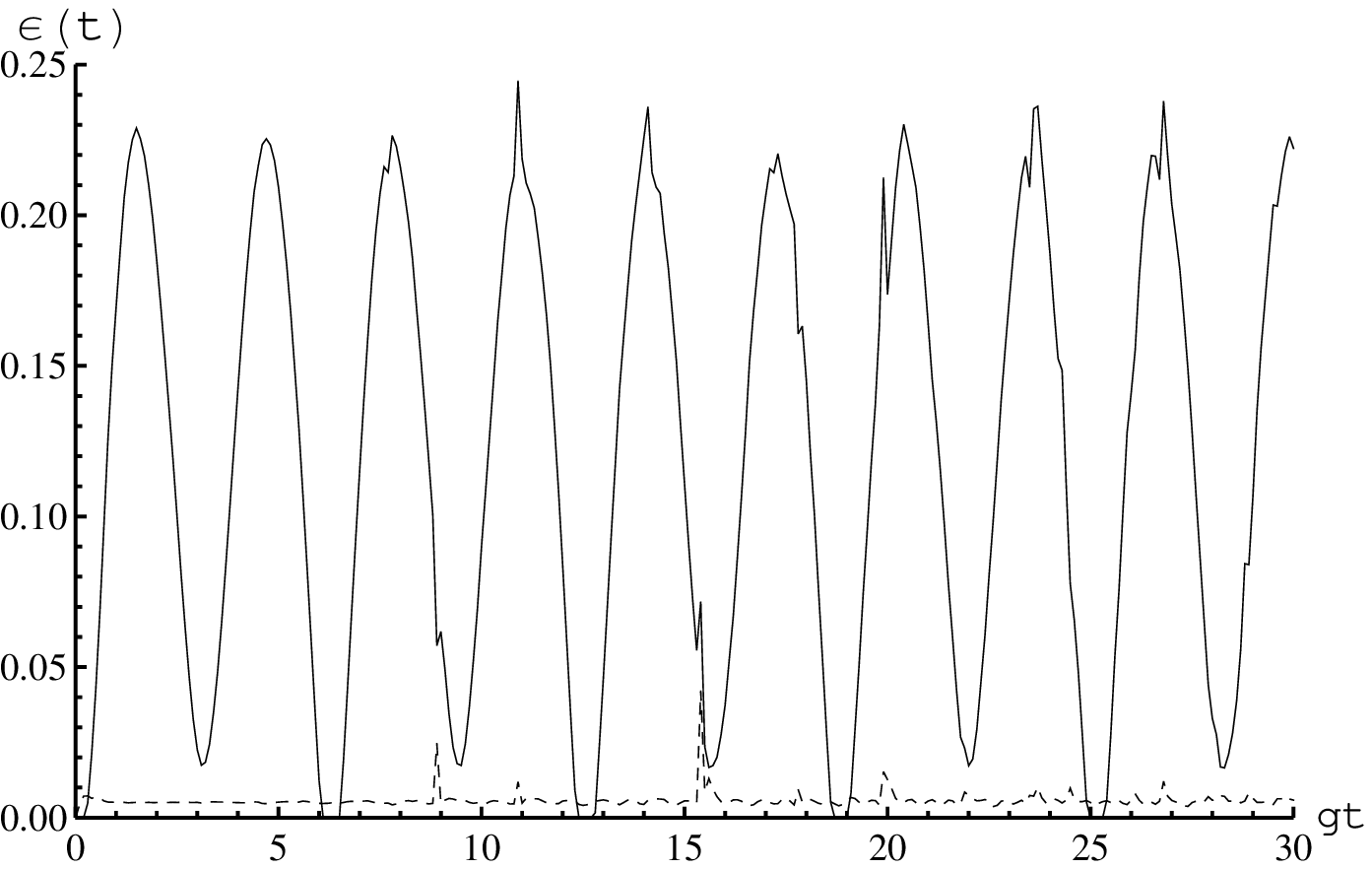}
\end{tabular}
 \caption{The negativity as a function of $gt$  for the model  with  $\bar n_1 = \bar n_2 =20 $: a)   $\alpha = 0.3$,   $|\Psi(0)\rangle _1= |+\rangle, |\Psi(0)\rangle_2 = |-\rangle$ (dashed), $|\Psi(0)\rangle_1=(1/\sqrt{2}(|+\rangle + |-\rangle), |\Psi(0)\rangle_2 = (1/\sqrt{2}(|+\rangle - |-\rangle) $ (solid), b) $\alpha=1$,  $|\Psi(0)\rangle _1= |+\rangle, |\Psi(0)\rangle_2 = |-\rangle$ (dashed), $|\Psi(0)\rangle_1=(1/\sqrt{2}(|+\rangle + |-\rangle), |\Psi(0)\rangle_2 = (1/\sqrt{2}(|+\rangle - |-\rangle) $ (solid).  }
\end{figure}

\section{Conclusion}
In this paper we have investigated the effect of the atomic coherence on the entanglement of two two-level identical atoms interacting with two-mode thermal field through a nondegenerate two-photon processes in the lossless cavity taking into account the dipole interaction. It is shown that the entanglement is dependent on the initial atomic state and strength  of the dipole interaction. The results also  show that the atom-atom entanglement can be controlled by changing the system parameters, such as relative phases, the amplitudes of the polarized atoms, the mean photon numbers of the two-mode thermal field, and the  strength  of the dipole interaction (or distance between atoms).  The have derived that the dipole-dipole interaction can produce the appreciable amount of  entanglement even in the case of large thermal noise if the atoms to be prepared in coherent states. In this paper we consider only one type of the atomic initial coherent states. The entanglement behavior  for initial atomic coherent states $|\Psi(0)\rangle = \cos \theta |+,-\rangle + \sin\theta |-,+\rangle$ and
 $|\Psi(0)\rangle = \cos \theta |+,+\rangle + \sin\theta |-,-\rangle$ will be considered in our following paper.

\section*{Acknowledgments}

Work is carried out with financial support of the Ministry of Education and Science of the Russian Federation (the state assignment N 2.2459.2011).

\section*{Appendix: Evident expressions for density matrix Elements in Eq. (6)}

$$\rho_{11}(t)=\sum\limits_{n_1=0}^{\infty} \sum\limits_{n_2=0}^{\infty} p_1(n_1) p_2(n_2)
\rho_{11}(0) \left ( \left(1+ 2 (n_1+1) (n_2+1) \frac{A_{n_1+1,n_2+1}}{\lambda_{n_1+1,n_2+1}}\right) \times \right.$$ $$\left. \times  \left(1+2
(n_1+1) (n_2+1)  \frac{A^*_{n_1+1,n_2+1}}{\lambda_{n_1+1,n_2+1}}\right)+
(\rho_{22}(0)+\rho_{32}(0)+\rho_{23}(0)+\rho_{33}(0)) n_1 n_2 \frac{|B_{n_1,n_2}|^2}{\theta_{n_1,n_2}^2}\right ) + $$ $$
+4 \sum\limits_{n_1=2}^{\infty} \sum\limits_{n_2=2}^{\infty} p_1(n_1) p_2(n_2)
 \rho_{44}(0) n_1 n_2 (n_1-1) (n_2-1) \frac{|A_{n_1-1,n_2-1}|^2}{\lambda_{n_1-1,n_2-1}^2};$$
$$\rho_{12}(t)=\sum\limits_{n_1=0}^{\infty} \sum\limits_{n_2=0}^{\infty} p_1(n_1) p_2(n_2)
(\rho_{12}(0) (U^*_{22})_{n_1,n_2} + \rho_{13}(0) (U^*_{22})_{n_1,n_2} ) \times $$ $$\times   \left(1+ 2 (n_1+1) (n_2+1) \frac{A_{n_1+1,n_2+1}}{\lambda_{n_1+1,n_2+1}}\right)  +$$
$$ + \sum\limits_{n_1=1}^{\infty} \sum\limits_{n_2=1}^{\infty} p_1(n_1) p_2(n_2) (\rho_{24}(0)+\rho_{34}(0)) n_1 n_2
\frac{B_{n_1,n_2}}{\theta_{n_1,n_2}} \frac{B^*_{n_1-1,n_2-1}}{\theta_{n_1-1,n_2-1}},  $$
$$\rho_{13}(t)=\sum\limits_{n_1=0}^{\infty} \sum\limits_{n_2=0}^{\infty} p_1(n_1) p_2(n_2)
(\rho_{12}(0) (U^*_{32})_{n_1,n_2} + \rho_{13}(0) (U^*_{33})_{n_1,n_2} ) \times $$ $$\times  \left(1+ 2 (n_1+1) (n_2+1) \frac{A_{n_1+1,n_2+1}}{\lambda_{n_1+1,n_2+1}}\right)  +$$
$$ + \sum\limits_{n_1=1}^{\infty} \sum\limits_{n_2=1}^{\infty} p_1(n_1) p_2(n_2) (\rho_{24}(0)+\rho_{34}(0)) n_1 n_2
\frac{B_{n_1,n_2}}{\theta_{n_1,n_2}} \frac{B^*_{n_1-1,n_2-1}}{\theta_{n_1-1,n_2-1}},  $$
$$\rho_{14}(t)=\sum\limits_{n_1=0}^{\infty} \sum\limits_{n_2=0}^{\infty} p_1(n_1) p_2(n_2)
\rho_{14} \left (\left(1+ 2 (n_1+1) (n_2+1) \frac{A_{n_1+1,n_2+1}}{\lambda_{n_1+1,n_2+1}}\right) \times \right.$$ $$\left. \times  \left(1+2
(n_1+1) (n_2+1)  \frac{A^*_{n_1-1,n_2-1}}{\lambda_{n_1-1,n_2-1}}\right)+ p_1(0)p_2(0) \rho_{14}(0) \left (1 + 2 \frac{A_{1,1}}{\lambda_{1,1}}\right )\right ) +$$
$$ + \sum\limits_{n_1=1}^{\infty} p_1(n_1) p_2(0) \rho_{14}\left (1 + 2 \frac{A_{n_1+1,1}}{\lambda_{n_1+1,1}}\right ) + \sum\limits_{n_2=1}^{\infty} p_1(0) p_2(n_2) \rho_{14}\left (1 + 2 \frac{A_{1,n_2+1}}{\lambda_{1+1,n_2+1}}\right )  $$
$$\rho_{22}(t)=\sum\limits_{n_1=0}^{\infty} \sum\limits_{n_2=0}^{\infty} p_1(n_1) p_2(n_2)
\left (\rho_{11}(0) (n_1+1)(n_2+1)\frac{|B_{n_1+1,n_2+1}|^2}{\theta_{n_1+1,n_2+1}^2} + \right.$$ $$ \left. + \rho_{22}(0) (U_{22})_{n_1,n_2}(U_{22})^*_{n_1,n_2}+  \rho_{32}(0) (U_{23})_{n_1,n_2}(U_{22})^*_{n_1,n_2} + \rho_{23}(0) (U_{22})_{n_1,n_2}(U_{23})^*_{n_1,n_2} + \right. $$
$$\left. + \rho_{33}(0) (U_{23})_{n_1,n_2}(U_{23})^*_{n_1,n_2} \right )
+ \sum\limits_{n_1=1}^{\infty} \sum\limits_{n_2=1}^{\infty} p_1(n_1) p_2(n_2)
\rho_{44}(0) n_1 n_2  \frac{|B_{n_1-1,n_2-1}|^2}{\theta_{n_1-1,n_2-1}^2}, $$
$$\rho_{23}(t)=\sum\limits_{n_1=0}^{\infty} \sum\limits_{n_2=0}^{\infty} p_1(n_1) p_2(n_2)
\left (\rho_{11}(0)(n_1+1)(n_2+1) \frac{|B_{n_1+1,n_2+1}|^2}{\theta_{n_1+1,n_2+1}^2} + \right .$$ $$\left. + \rho_{22}(0) (U_{22})_{n_1,n_2}(U_{32})^*_{n_1,n_2}
+\rho_{32}(0) (U_{23})_{n_1,n_2}(U_{32})^*_{n_1,n_2} + \rho_{23}(0) (U_{22})_{n_1,n_2}(U_{33})^*_{n_1,n_2} + \right. $$ $$ \left. + \rho_{33}(0) (U_{23})_{n_1,n_2}(U_{33})^*_{n_1,n_2} \right ) +
\sum\limits_{n_1=0}^{\infty} \sum\limits_{n_2=0}^{\infty} p_1(n_1) p_2(n_2)
 n_1 n_2  \rho_{44}(0)   \frac{B_{n_1,n_2}}{\theta_{n_1,n_2}} ,
$$
$$\rho_{24}(t)
 \sum\limits_{n_1=0}^{\infty} \sum\limits_{n_2=0}^{\infty} p_1(n_1) p_2(n_2)
 (n_1+1) (n_2+1)  (\rho_{12}(0)  + \rho_{13}(0))  \frac{B^*_{n_1,n_2}}{\theta_{n_1,n_2}} \frac{B_{n_1+1,n_2+1}}{\theta_{n_1+1,n_2+1}}  +$$
$$  + \sum\limits_{n_1=1}^{\infty}  p_1(n_1) p_2(0) (\rho_{24}(0)(U_{22})_{n_1,0}+\rho_{34}(0)(U_{23})_{n_1,0}) + $$ $$ +
  \sum\limits_{n_2=1}^{\infty}  p_1(0) p_2(n_2) (\rho_{24}(0)(U_{22})_{0,n_2}+\rho_{34}(0)(U_{23})_{0,n_2}) +$$ $$  + p_1(0) p_2(0) (\rho_{24}(0)(U_{22})_{0,0}+\rho_{34}(0)(U_{23})_{0,0}) +$$
$$
+ \sum\limits_{n_1=1}^{\infty} \sum\limits_{n_2=1}^{\infty} p_1(n_1) p_2(n_2)  (\rho_{24}(0)(U_{22})_{n_1,n_2}+\rho_{34}(0)(U_{23})_{n_1,n_2}) \times$$
$$\times \left (1 + 2 n_1 n_2 \frac{A^*_{n_1-1,n_2-1}}{\lambda_{n_1-1,n_2-1}}\right ), $$
$$\rho_{33}(t)
 \sum\limits_{n_1=0}^{\infty} \sum\limits_{n_2=0}^{\infty} p_1(n_1) p_2(n_2)\rho_{11}(0) \left (
  (n_1+1) (n_2+1)    \frac{|B_{n_1+1,n_2+1}|}{\theta_{n_1+1,n_2+1}} + \right. $$ $$ \left. +
  \rho_{23}(0)(U_{32})_{n_1,n_2}(U^*_{33})_{n_1,n_2}+ \rho_{32}(0)(U_{33})_{n_1,n_2}(U^*_{32})_{n_1,n_2} +\right. $$
$$\left.+  \rho_{22}(0)(U_{32})_{n_1,n_2}(U^*_{32})_{n_1,n_2}+ \rho_{33}(0)(U_{33})_{n_1,n_2}(U^*_{33})_{n_1,n_2} \right )+$$
$$+ \sum\limits_{n_1=1}^{\infty} \sum\limits_{n_2=1}^{\infty} p_1(n_1) p_2(n_2)\rho_{44}(0) n_1 n_2    \frac{|B_{n_1-1,n_2-1}|}{\theta_{n_1-1,n_2-1}}, $$
$$\rho_{34}(t)
 \sum\limits_{n_1=0}^{\infty} \sum\limits_{n_2=0}^{\infty} p_1(n_1) p_2(n_2)\rho_{11}(0)
  (n_1+1) (n_2+1)    \frac{B^*_{n_1,n_2}}{\theta_{n_1,n_2}} \frac{B_{n_1+1,n_2+1}}{\theta_{n_1+1,n_2+1}} +$$
$$+  \sum\limits_{n_1=1}^{\infty} p1(n_1) p2(0)(\rho_{24} (U_{32})_{n_1,0} + \rho_{34} (U_{33})_{n_1,0} +$$
  $$+  \sum\limits_{n_2=1}^{\infty} p1(0) p2(n_2)(\rho_{24} (U_{32})_{0,n_2} + \rho_{34} (U_{33})_{0,n_2} +$$
  $$+   p1(0) p2(0)(\rho_{24} (U_{32})_{0,0} + \rho_{34} (U_{33})_{0,0} +$$
  $$ + \sum\limits_{n_1=1}^{\infty} \sum\limits_{n_2=1}^{\infty} p_1(n_1) p_2(n_2)(\rho_{24}(0) (U_{32})_{n_1,n_2} +
       \rho_{34}(0) (U_{33})_{n_1,n_2}) \times $$ $$ \times \left (1 + 2 n_1 n_2    \frac{A^*_{n_1-1,n_2-1}}{\lambda_{n_1-1,n_2-1}}\right ).$$


\section*{References}
\begin{thebibliography}{99}
\bibitem{Nielsen} M.A.Nielsen M.A., I.L. Chuang, Quantum Computation and Quantum Information,
Cambrige: Cambrige University Press, 2000.
\bibitem{Plenio} M.B. Plenio, S.F. Huelda, A. Beige, P.L. Knight, Phys. Rev. A. {\bf 59}, 2468-2475 (1999).
\bibitem{Bose} S. Bose S., I. Fruentes-Guridi, P.L. Knight, V. Vedral, Phys.Rev.Lett. {\bf 87}, 050401(1-4)  (2001).
\bibitem{Kim} M.S. Kim, J. Lee, D. Ahn, P.L. Knight, Phys. Rev. A. {\bf 65}, 040101(1-4)  (2002).
\bibitem{Zhou1} L. Zhou, H.S. Song, J. Optics B: Quantum Semiclass. Opt.  {\bf 6}, 378 - 384 (2004).
\bibitem{Zhou2} L. Zhou, H.S. Song, J. Optics B: Quantum Semiclass. Opt.  {\bf 4}, 425-429 (2002).
\bibitem{Bash1} Bashkirov E.K.,  Laser Phys. Lett.    {\bf 3},    145-150 ( 2006).
\bibitem{Aguiar} Aguiar L.S.,  Munhoz P.P.,   Vidiella-Barranco A,   Roversi J.A.,  J. Opt.   {\bf B7} S769-771 (2005).
\bibitem{Liao} Liao X-P.,   Fang M-F.,   Cai J-W.,   Zheng X-J.,  Chin. Phys.  {\bf B17} 2137-2142 (2008).
\bibitem{Bash2}	 Bashkirov E.K.,  Stupatskaya M.P.,  Laser Phys.    {19} 525-530 (2009).
 \bibitem{Hu1} Hu Y.-H., Fang M.-F., Wu Q.,  Chin. Phys.  {\bf 16} 2407-2414 (2007).
\bibitem{Hu2} Hu Y.-H., Fang M.-F., Jiang C.-L., Zeng K.,   Chin. Phys.   {\bf 17}1784-1790 (2008).
\bibitem{Hu3} Hu Y.-H., Fang M.-F.,  Comm. Theor. Phys.  {\bf 54} 421-426  (2010).
\bibitem{Peres}	Peres A.,   Phys. Rev. Lett.  {\bf 77}   1413 - 1415 (1996).
\bibitem{Horod}	Horodecki R., 	Horodecki M., 	Horodecki P.,  Phys. Lett. {\bf A223}  (333-339) (1996).

\end{thebibliography}
\end{document}